\documentclass[a4paper]{article}
\usepackage[table]{xcolor}
\usepackage{spconf,graphicx}
\usepackage[hyphens]{url}
\usepackage[]{hyperref}
\usepackage{tikz}
\usepackage{booktabs}
\usepackage{multirow}

\hypersetup{breaklinks=true}
\usepackage{INTERSPEECH2022}

\title{Coswara: A website application enabling COVID-19 screening by analysing respiratory sound samples and health symptoms}
 \name{Debarpan Bhattacharya$^1$,
 Debottam Dutta$^1$, Neeraj Kumar Sharma$^2$, Srikanth Raj Chetupalli$^2$, Pravin Mote$^1$, Sriram Ganapathy$^1$, Chandrakiran C$^3$, Sahiti Nori$^3$, Suhail K K$^3$, Sadhana Gonuguntla$^4$, Murali Alagesan$^5$
 }
\address{
   $^1$LEAP lab, Indian Institute of Science, Bangalore, India,
   $^2$Erlangen, Germany,
   $^3$Ramaiah Medical College Hospital, Bangalore, India,
   $^4$General Hospital, Hoskote, Bangalore, India,
   $^5$PSG Institute of Medical Sciences and Research, India
   }
\email{sriramg@iisc.ac.in}
\ninept 
\begin{document}
\maketitle
\begin{abstract}
The COVID-19 pandemic has accelerated research on design of alternative, quick and effective COVID-19 diagnosis approaches. In this paper, we describe the Coswara tool, a website application designed to enable COVID-19 detection by analysing respiratory sound samples and health symptoms. 
A user using this service can log into a website using any device connected to the internet, provide there current health symptom information and record few sound sampled corresponding to breathing, cough, and speech. Within a minute of analysis of this information on a cloud server the website tool will output a COVID-19 probability score to the user. As the COVID-19 pandemic continues to demand massive and scalable population level testing, we hypothesize that the proposed tool provides a potential solution towards this.

\end{abstract}
\noindent\textbf{Index Terms}: COVID-19, web application, cough, breathing, vowel, counting and speech.
\begin{figure*}[t!]
    \centering
\begin{tikzpicture}
\node[inner sep=0pt,anchor=center,draw=black, line width=0.5mm] (main_fig) at (2,10) 
{\includegraphics[width=3.3cm,height=7cm]{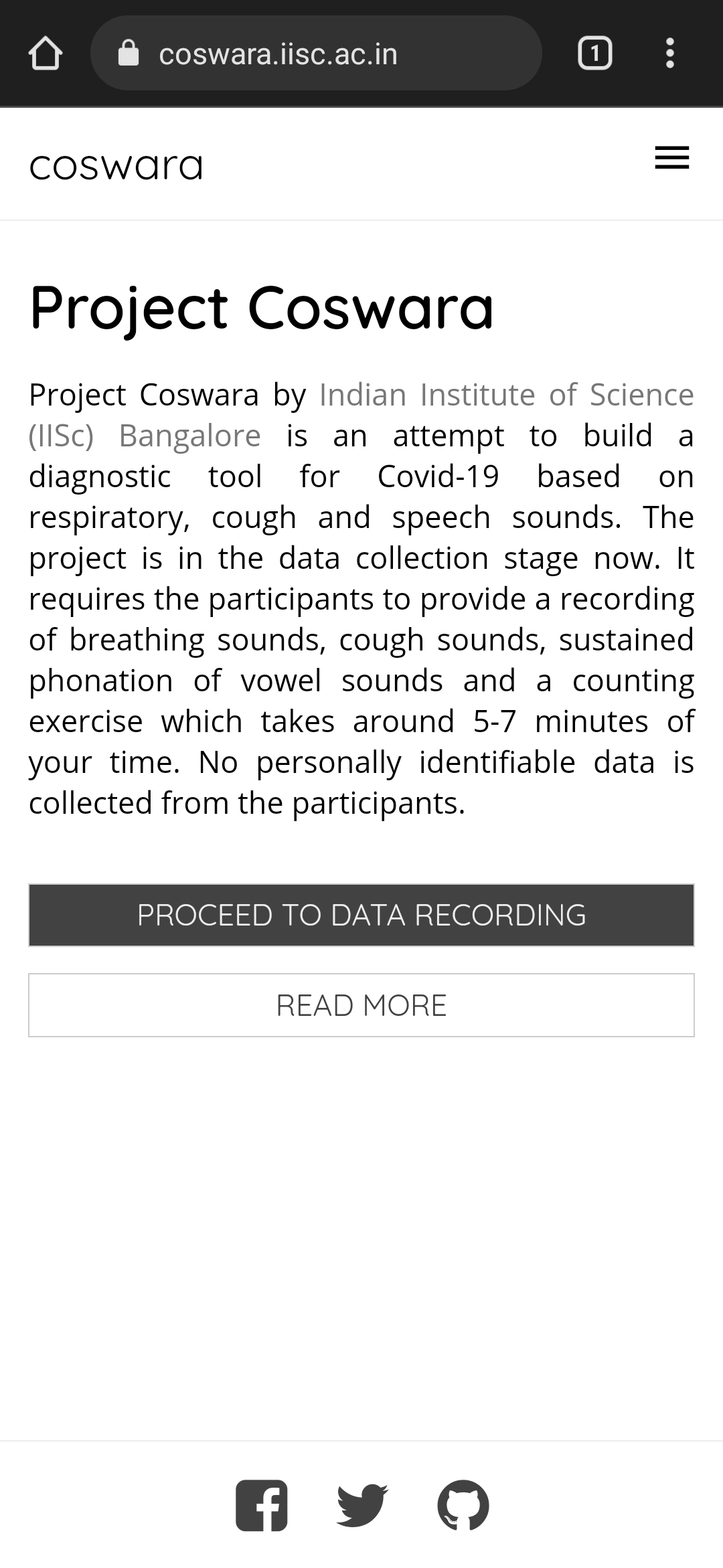}};

\node[inner sep=0pt,anchor=center,draw=black, line width=0.5mm] (main_fig) at (5.5,10) 
{\includegraphics[width=3.3cm,height=7cm]{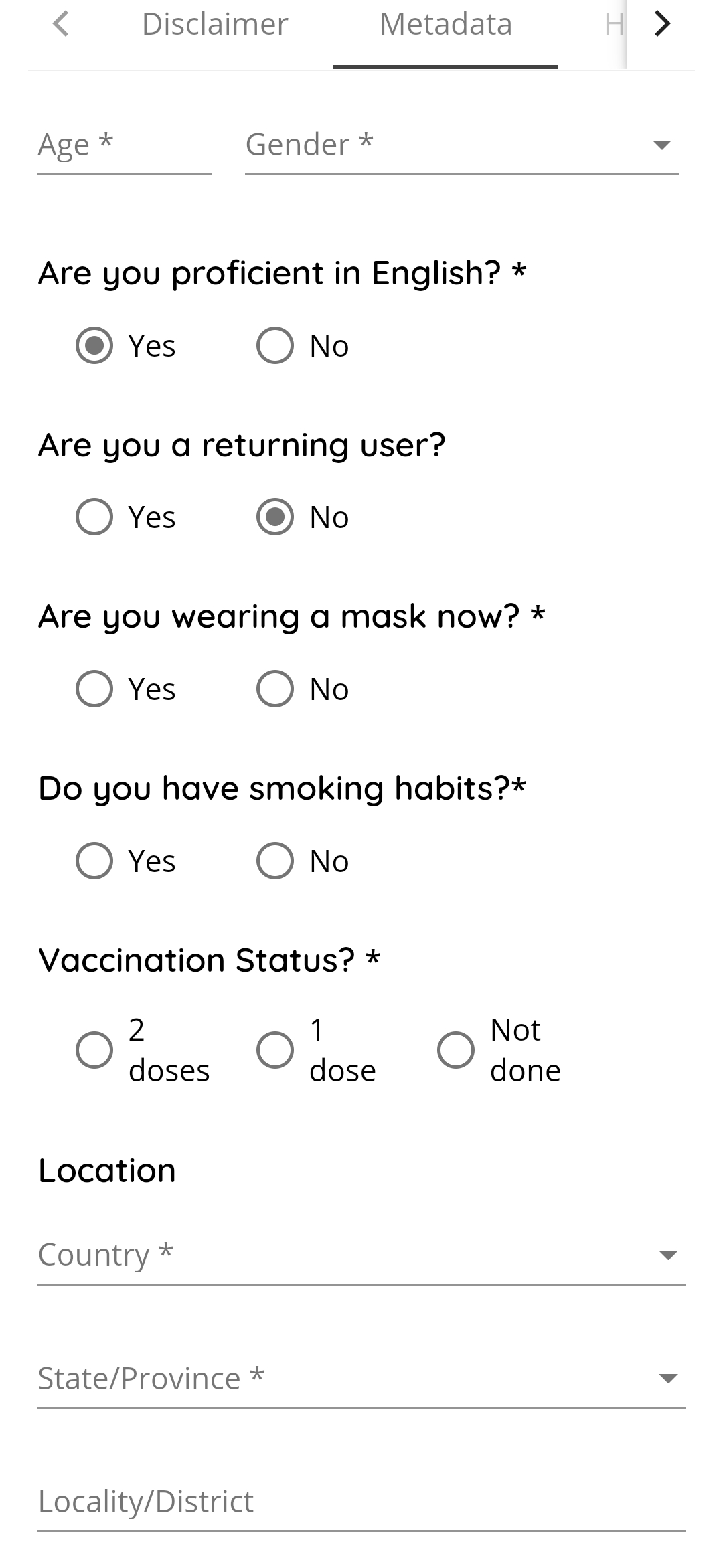}};

\node[inner sep=0pt,anchor=center,draw=black, line width=0.5mm] (main_fig) at (9,10) 
{\includegraphics[width=3.3cm,height=7cm]{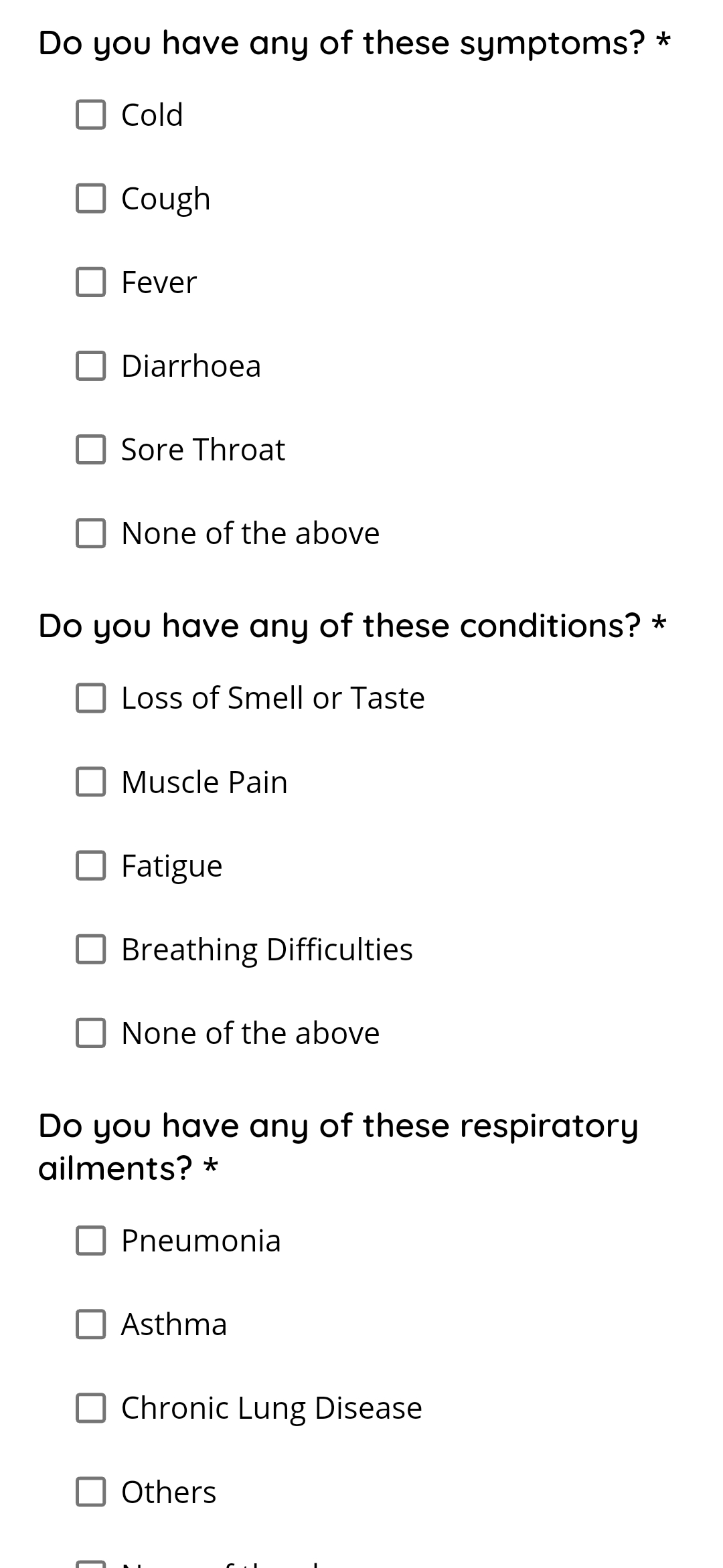}};

\node[inner sep=0pt,anchor=center,draw=black, line width=0.5mm] (main_fig) at (12.5,10) 
{\includegraphics[width=3.3cm,height=7cm]{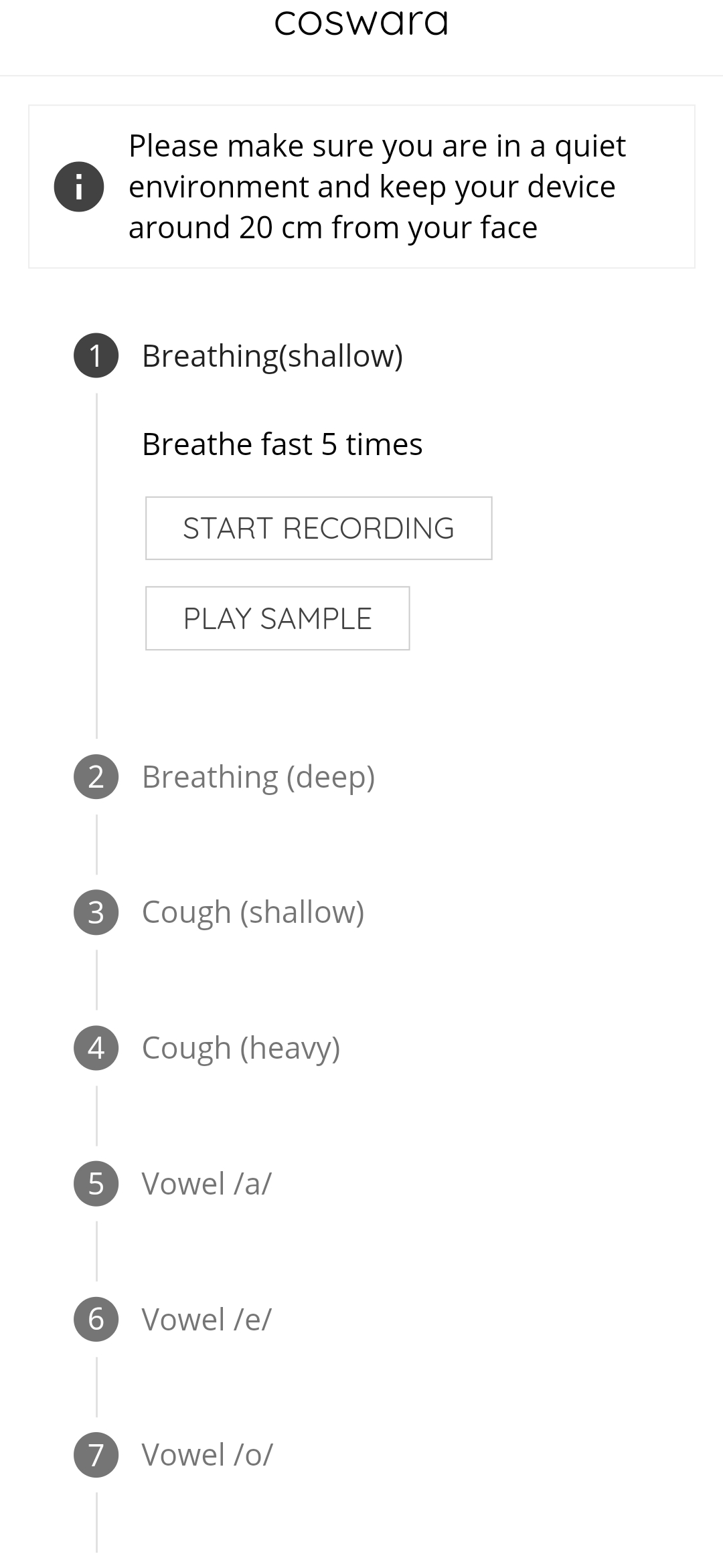}};

\node[inner sep=0pt,anchor=center,draw=black, line width=0.5mm] (main_fig) at (16,10) 
{\includegraphics[width=3.3cm,height=7cm]{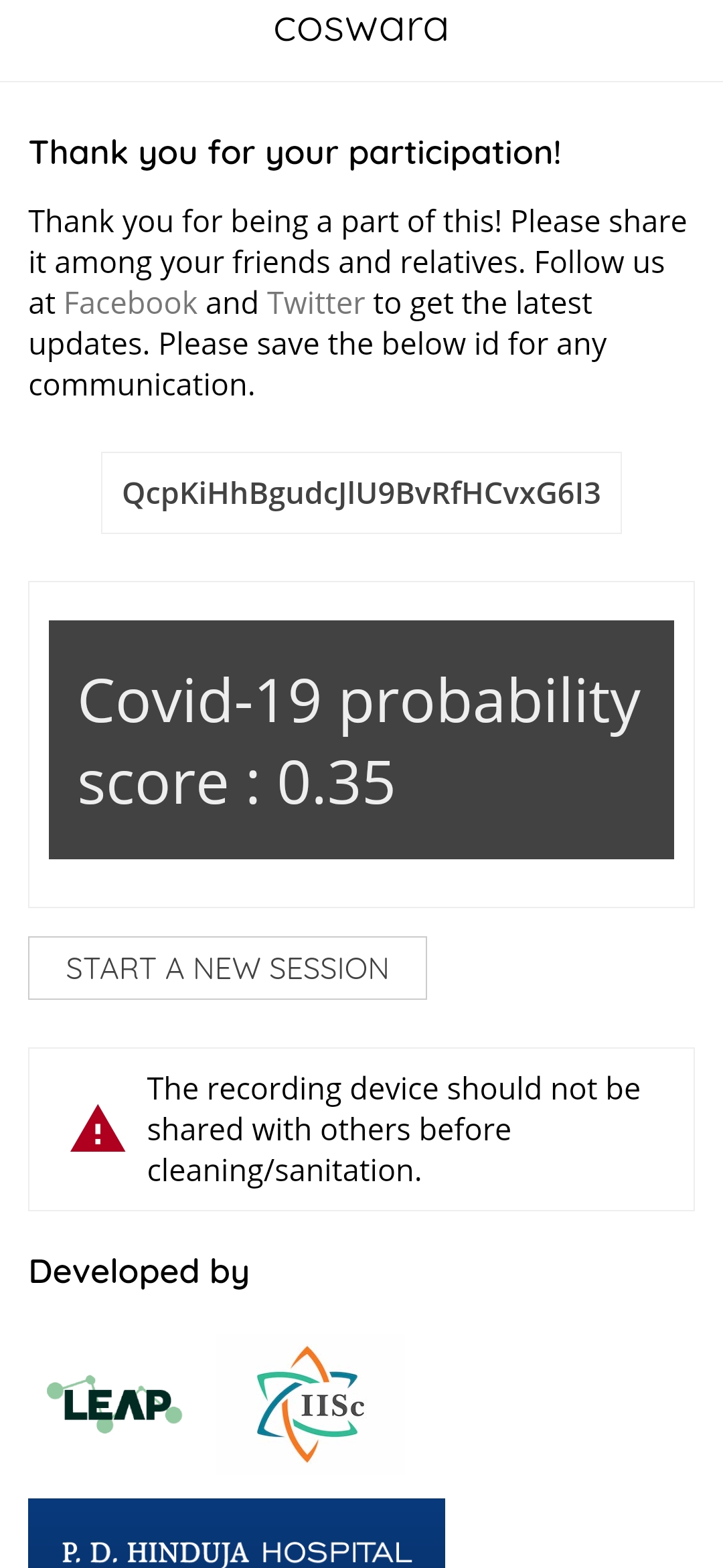}};

\node[font=\fontsize{8}{6}\selectfont,rotate=0,anchor=center] at (2,6) (l1) {(a)};
\node[font=\fontsize{8}{6}\selectfont,rotate=0,anchor=center] at (5.5,6) (l1) {(b)};
\node[font=\fontsize{8}{6}\selectfont,rotate=0,anchor=center] at (9,6) (l1) {(c)};
\node[font=\fontsize{8}{6}\selectfont,rotate=0,anchor=center] at (12.5,6) (l1) {(d)};
\node[font=\fontsize{8}{6}\selectfont,rotate=0,anchor=center] at (16,6) (l1) {(e)};

\end{tikzpicture}
    \vspace{-0.25in}
    \caption{The Coswara website: (a) welcome page, (b) user metadata collection, (c) health symptoms collection, (d) sound sample collection, (e) displaying COVID-19 probability score.}
    \label{fig:app_ss}
\vspace{-0.25in}
\end{figure*}
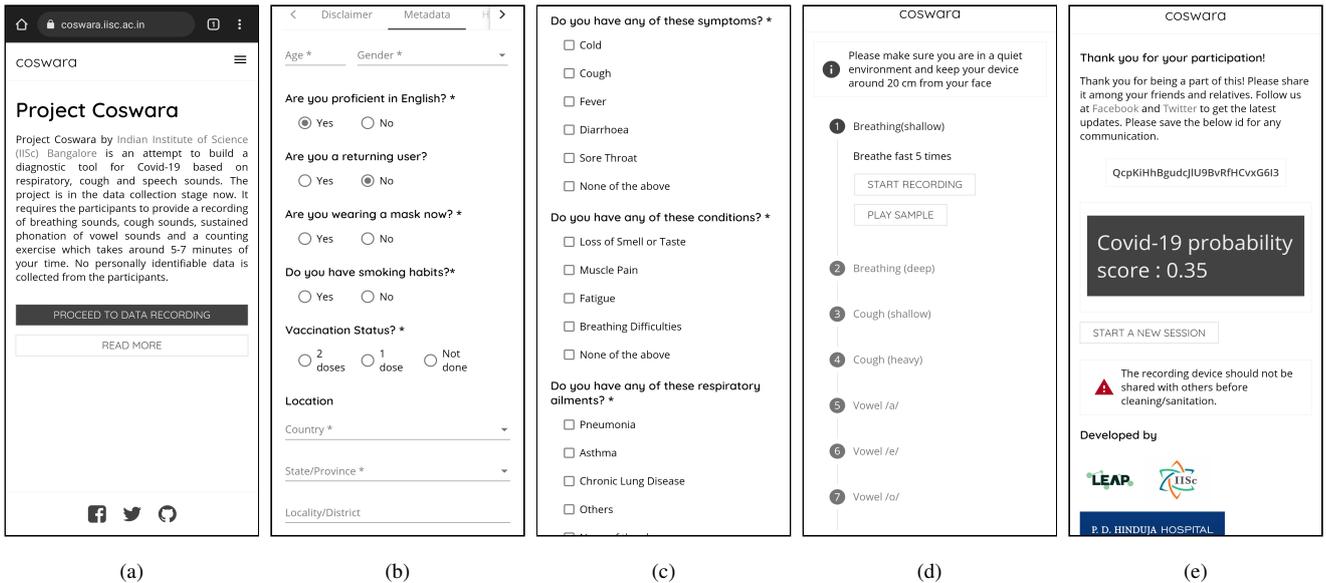

\section{Introduction}
The COVID-19 RT-PCR tests, a current gold standard for diagnosis, are based on molecular testing and detect the amino-acid sequences unique to SARS-CoV-2 in the swab samples. Usage of these tests for population level testing suffers from few limitations, namely, i) high cost of RT-PCR chemical reagents and facility, ii) need for expert supervision, iii) variable turnaround time from sample collection to results (hours to days), and iv) lack of physical distancing during sample collection. Recognizing this, the WHO's blueprint on COVID-19 diagnostic tests highlights the urgent need for developing point-of-care tests (POCTs) \cite{who_poct}. Our demonstration will share a COVID-19 screening methodology which is based on analyzing the respiratory sound samples and health symptoms of a person. It aims at providing improvement with respect to time, cost, and physical distancing, and without trading-off the detection performance.

\begin{figure}[t]
    \centering
    \input{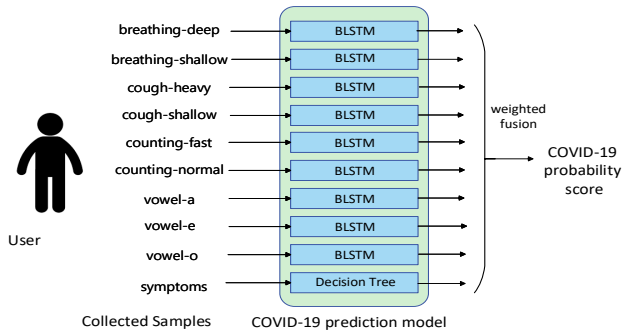}
     \vspace{-0.05in}
    \caption{Proposed COVID-19 screening methodology.}
    \label{fig:app_score}
 \vspace{-0.05in}
\end{figure}
\begin{table}[]
\centering
\rowcolors{2}{white}{gray!10}
\begin{tabular}{@{}ll@{}}
\toprule
\textbf{Category} &   \multicolumn{1}{c}{\textbf{Details}} \\ \midrule
 \begin{tabular}[c]{@{}l@{}}breathing-deep \\ breathing-shallow\end{tabular} &   \begin{tabular}[c]{@{}l@{}}Few respiration cycles in deep and\\ shallow manner\end{tabular}  \\
 \begin{tabular}[c]{@{}l@{}}cough-heavy \\ cough-shallow\end{tabular} &   \begin{tabular}[c]{@{}l@{}}Few cycles of coughing in heavy and\\ shallow manner\end{tabular} \\
 \begin{tabular}[c]{@{}l@{}}counting-fast \\ counting-normal\end{tabular} &  \begin{tabular}[c]{@{}l@{}}speech sound of counting 1 to 20 in\\ fast and normal pace\end{tabular} \\
 \begin{tabular}[c]{@{}l@{}}vowel-a, vowel-e, \\ vowel-o\end{tabular} &  \begin{tabular}[c]{@{}l@{}}sustained  phonation of vowels: $[$u] in \\boot, $[$i] in beet, [\ae] in bat\end{tabular} \\
 \begin{tabular}[c]{@{}l@{}}health symptoms\end{tabular} &  \begin{tabular}[c]{@{}l@{}}symptoms of the patient for example,\\ cold, fever, breathing difficulty etc.
 \end{tabular} \\\bottomrule
\end{tabular}
\caption{Description of nine sound categories and symptoms data collected by the website application.}
\label{table:data_details}
    \vspace{-0.2in}
\end{table}

\section{Coswara Tool}
The Coswara COVID-19 screening tool is a website application. It functions as follows. After a user logs into the website \url{http://coswara.iisc.ac.in/} and opts for a language of their choice, it request them to contribute their current health status information. For this they are provided with a a short questionnaire web page (shown in Figure~\ref{fig:app_ss}(c)) which collects information such as pre-existing health conditions, and current health symptoms (including cough, cold, diarrhoea, muscle pain, and breathing difficulty). Subsequently, it requests the user to contribute their sound samples using the device microphone (shown in Figure~\ref{fig:app_ss}(d)). Nine kinds of respiratory sound samples, listed in Table \ref{table:data_details}, are collected. On pressing the submit button all this information is sent to a cloud server, analyzed and in a few seconds the user is provided with a COVID-19 probability score indicating how likely they have a COVID-19 infection.
The tool is not yet approved by the healthcare regulatory authority of India. Hence, the use of the designed tool is subject to agreeing with the terms and conditions dictated on the web page.

\subsection{Methodology}
The tool is based on using a binary classification approach to detect individuals infected by COVID-19 disease. It assumes that the respiratory sound samples, originating from the lungs, are significantly different between healthy individuals and those infected with COVID-19 disease. This hypothesis is validated in multiple studies by different research groups \cite{imran2020ai4, brown2020exploring, xia2021covid, chetupalli2021multi, sharma2022towards}. Further, it also makes use of the health symptom information to aid in the prediction of how likely a person has COVID-19. Utilizing such information for quick screening of COVID-19 as also been suggested in \cite{zoabi2021machine}.

The designed classifiers are trained using the publicly available Coswara dataset \cite{sharma2020coswara}. From sound samples, the mel-spectrogram acoustic features are extracted and analyzed using a bi-directional LSTM (BLSTM) classifier. This classifier gave significantly better than chance level in the task of COVID-19 detection from respiratory sound samples ($>80\%$ AUC). The classifier has also served as baseline in the Second DiCOVA challenge\cite{sharma2021second}. The BLSTM based classifiers are tuned separately for each of the nine sound categories. A decision tree classifier is used for analysis of the health symptom information. The final model is a fusion of COVID-19 probability scores obtained from the ten models ($9$ sound category models and $1$ symptom model). An illustration is provided in Figure~\ref{fig:app_score}.

\section{Conclusion}
The Coswara website application demonstrates a journey from defining a new problem, that is, COVID-19 detection from sound samples, creating and curating a dataset to help analyze the problem, and subsequently, design solutions for it. We consider that demonstrating the tool will help us share the journey and also the open challenges with other researchers. The designed tool shows potential for making a societal impact, and a need for deepening understanding of respiratory acoustics.
\section{Acknowledgement}
The authors would like to thank the Department of Science and Technology (DST), Government of India, for funding the Coswara Project through the RAKSHAK program.
\bibliographystyle{IEEEtran}

\bibliography{mybib}


\end{document}